# La Anaconda y el Dragón – Sobre la estructura de la colaboración científica institucional entre América Latina y China

*Julián D. Cortés*

*Universidad del Rosario, Colombia*

*Fudan University, China*

*E-mail: julian.cortess@urosario.edu.co*

**Hallazgos**

- Las coautorías institucionales entre ALC y China fueron fertilizada mediante su participación en proyectos globales relacionados con liderazgo.
- Las instituciones de cada región divergieron de su posición inicial en la red de coautorías: ALC a la periferia y China al centro.
- Las comunidades institucionales se han compactado y han disminuido sus enlaces con comunidades externas.
- Se incrementó el ingreso de nuevas instituciones, especialmente en el último sexenio.
- Instituciones chinas exhibieron en su momento un protagonismo como puentes sociales. Lugar que posteriormente ocupó una institución colombiana.
- Australia se consolida como la geografía que alberga las instituciones con mayor intermediación.

**Palabras clave:** América Latina; China; Coautoría; Análisis de Redes Sociales; Economía y Negocios.

## 1    Introducción

En 1565 desembarcó el Galeón de Manila en Acapulco, México. Así inició la migración de la población china a América Latina y el Caribe (ALC) (Connell & Cornejo Bustamante, 1992). La anaconda y el dragón comenzaron a entrelazarse. Más de 500 años después, se estimó que ambas regiones crecerían más rápido que las economías (post)industrializadas gracias al resultado de intercambios comerciales y tecnológicos en un marco de cooperación Sur-Sur (CEPAL, 2012). Sin pasar por alto que ambas regiones enfrentan una constelación de problemas sociales y medio ambientales por aplacar (CEPAL, 2019; Rohde & Muller, 2015).

ALC y China han afianzado sus lazos para la colaboración científica (e.g., Consorcio de Universidades Fudan América Latica-FLAUC (FLAUC, s/f). Este último factor ya es un común denominador en la generación moderna de conocimiento (Adams, 2012; Wuchty et al., 2007). Como resultado, el porcentaje de los artículos de investigación en el que participan coautores de diversos países se duplicó en los últimos 20 años (Wagner et al., 2015). La colaboración científica incide en una mayor productividad, efectividad en el uso de recursos, intersección e intercambio de competencias y conocimiento especializados, y mayor impacto (Adams et al., 2019; Li et al., 2013). A pesar de ello, las dinámicas de colaboración no son homogéneas (e.g., la colaboración interna entre Brasil, India, Rusia, China y Sudáfrica – BIRCS es menor que la que desarrolla con otros países) (Finardi & Buratti, 2016).

Así y todo, las investigaciones sobre colaboración científica entre ALC y China no se han desarrollado con la misma diligencia que, por ejemplo, la agenda sobre los análisis económicos o geopolíticos (Gonzalez-Vicente, 2012; Lemarchand, 2012; Lin & Treichel, 2012; Mesquita Moreira, 2007; Myers & Wise, 2016). En consecuencia, esta ponencia presenta evidencia longitudinal sobre la colaboración científica institucional entre ALC y China. Los resultados podrían ser de utilidad para investigadores, centros de investigación, incluso, unidades de relaciones interinstitucionales y de cooperación internacional de universidades y gobiernos locales o nacionales mediante la elaboración de una cartografía científica que identifique los actores estratégicos que intervienen en la producción y colaboración científica entre ALC y China.

Aspectos relacionados con áreas de investigación, desempeño e impacto, se publicarán próximamente, luego no hacen parte de este estudio (Cortés-Sánchez, 2022). En las tres secciones siguientes, se abordará la metodología, los resultados y su discusión, y se definirán las conclusiones.

## 2    Metodología

Se escogió la base de datos bibliográfica Scopus (2020) por dos razones: una mayor participación de investigadores de países en vías de desarrollo en ciencias sociales y una mayor cobertura neta en el número de revistas a comparación de Web of Science (WoS) (Baas et al., 2020; Mongeon & Paul-Hus, 2016; Scopus, 2019). La selección de documentos se restringió a artículos de investigación en áreas relacionadas con negocios y economía (i.e., negocios, administración y contabilidad; ciencias de la decisión; y economía, econometría y finanzas) por tres factores: la similitud en las dinámicas de producción y coautoría entre las áreas seleccionadas, un creciente interés de estas áreas −negocios/administración especialmente− por la publicación en revistas indexadas en lugar de libros, y el interés particular de escuelas de negocios y economía, gobiernos y agencias de cooperación de ambas regiones para la difusión de conocimiento con la participación directa o indirecta del sector privado (Acedo et al., 2006; Adams et al., 2019; Cortés-Sánchez, 2019; Gingras, 2014). El período de tiempo postulado (1996-2019) se sustenta en la fidelidad de la inclusión de publicaciones después de 1996 y la publicación finalizada de los documentos al menos un año previo a la fecha de elaboración este estudio (Ioannidis et al., 2016). Este período se dividirá en cuatro segmentos: 1996-2001, 2002-2007, 2008-2013, y 2014-2019.

Se implementará el análisis de redes a las coautorías usando el paquete de R bibliometrix, y Gephi (Aria & Cuccurullo, 2017; Bastian et al., 2009; R Core Team, 2014). Se computarán para cada red dos métricas macro, una meso, y dos micro. Las métricas macro son densidad y longitud media de camino (LMC), la meso sobre modularidad, y las micro grado e intermediación (Scott, 1988). La densidad indica la proporción en la que la red se encuentra conectada en función del número de nodos (i.e., instituciones) y enlaces (i.e., artículo en donde figuran al menos dos instituciones de ALC y China) (i.e., densidad igual a uno indica que todos los nodos están conectados entre sí) y la LMC camino describe qué tan larga es una cadena de interacciones entre dos nodos de la red (i.e., longitud media de camino igual a dos indica que la universidad $X$ en Uruguay está a dos instituciones en el trayecto de una colaboración potencial con la universidad $Y$ en China) (Brandes, 2001; Iacobucci et al., 2018). La modularidad indica si la red tiene una estructura de comunidad (i.e., un mayor índice de modularidad quiere decir que las comunidades están fuertemente conectadas entre sí y débilmente con otras comunidades) (Blondel et al., 2008; Traag et al., 2019). Y el grado y la intermediación, revelan el número de enlaces que estableció cada institución y la capacidad de un nodo para interconectar múltiples comunidades, respectivamente (Opsahl et al., 2010) −una suerte de "puente social" entre comunidades.

## 3    Resultados y discusión

Las instituciones de ALC y China han publicado 302 artículos en coautoría, sobrepasando así el umbral para un estudio exploratorio de un cuerpo de conocimiento (i.e., ≈80 documentos) (Desrochers et al., 2016). Los artículos fueron elaborados por 1,876 autores exhibiendo un incremento anual de 16.4%. Únicamente seis artículos fueron publicados por un solo autor, seguramente, con múltiple afiliación institucional. El promedio de coautores por documento fue de ≈8. La Figura 1 presenta la densidad acumulada de publicaciones anual, con un incremento notable en los dos últimos años (i.e., 104 artículos) comparado con dos artículos publicados en 1997. No se publicaron artículos en 1996.

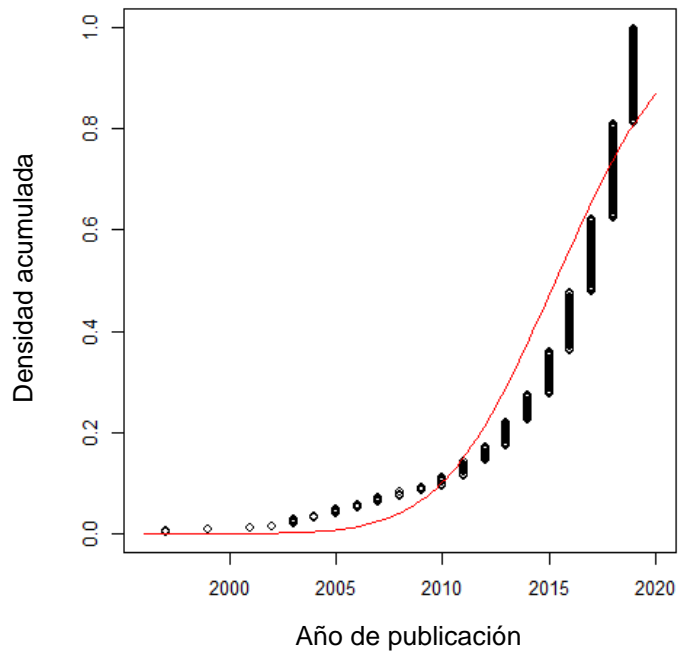

**Figura 1 Distribución y densidad de los artículos publicados 1997-2019. Fuente: el autor con base en Scopus (2019).**

La Tabla 1 presenta las métricas macro, meso y micro de las redes de coautoría. En una observación general (Figura 3), las instituciones de LAC se encontraban en una posición central a finales del siglo pasado, para irse desplazando paulatinamente a la periferia. El punto de partida de las instituciones de China es similar al de ALC, no obstante, con el transcurso del tiempo se fueron consolidando en el área central, entrelazándose así con instituciones del Norte Global y otros países de Asia. También se evidencia que las instituciones ALC y China han estado dentro del componente principal (i.e., la comunidad que agrupa el mayor porcentaje de nodos) durante todos los periodos.

Las métricas macro indican un incremento consistente en el número total de instituciones durante el periodo 2002-2019. En 2008-2013 el número total (35) y el incremento 2002-2013 de instituciones de ALC y China fueron similares: 118%+ y 133%+, respectivamente. En todo caso, no hay comparación con el incremento total de ALC, y de ambas regiones en la red de 2014-2019, lo que es apoyado por el incremento total de publicaciones (Figura 1). La progresiva disminución de la densidad ilustra que el paso al que se incrementa la participación de instituciones es mayor al número de enlaces que se han creado. En mayor detalle, mientras el LMC en 1997-2001 era de 1.4 (i.e., cualquier institución requería en promedio de una institución mediadora para conectarse con cualquier otra institución) en 2014-2019 fue de 3.5. El incremento consistente de la modularidad indica que con el paso del tiempo las redes de coautoría han forjado comunidades cada vez más interconectadas entre sí, aunque débilmente conectadas con comunidades externas (i.e., las comunidades institucionales de investigación son cada vez más definidas y difíciles de penetrar).

**Tabla 1 Métricas macro, meso, y micro de las redes de coautoría 1997-2019.**

| Métricas | 1997-2001 | 2002-2007 | Δ% | 2008-2013 | Δ% | 2014-2019 | Δ% |
|---|---|---|---|---|---|---|---|
| Total nodos | 130 | 128 | -1,5% | 288 | 125,0% | 901 | 212,8% |
| Nodos ALC | 18 | 16 | -11,1% | 35 | 118,8% | 105 | 200,0% |
| Nodos China | 22 | 15 | -31,8% | 35 | 133,3% | 91 | 160,0% |
| Densidad | 0,63 | 0,18 | -71,4% | 0,08 | -55,6% | 0,02 | -75,0% |
| LMC | 1,4 | 2 | 42,9% | 2,8 | 40,0% | 3,5 | 25,0% |
| Modularidad | 0,16 | 0,59 | 268,8% | 0,7 | 18,6% | 0,75 | 7,1% |
| Componente principal | 48.4% | 25.7% | -22.7 puntos % | 17.7% | -8 puntos % | 10.6% | -7.1 puntos % |
| Intermediación top-5 | U. Brasilia (0.015) | Babeş-Bolyai (0.23) | | U. Philippines (0.22) | | Aarhus U. (0.07) | |
| | Babeş-Bolyai U. (0.015) | Sultan Qaboos U. (0.11)** | | Victoria U. Wellington (0.16) | | U. New South Wales (0.07) | |
| | Sultan Qaboos U. (0.014)** | U. Ljubljana (0.11)** | | U. California (0.14) | | U. Gothenburg (0.06) | |
| | U. Ljubljana (0.014)** | Lignan U. (0.09) | | U. Texas (0.09) | | U. Andes (0.058) | |
| | Beijing U. Aero./Astronautics (0.013) | Capital Normal U. (0.0003) | | Monash U. (0.049) | | U. Queensland (0.056) | |

**Fuente: el autor con base en Scopus (2019) y procesado con bibliometrix (Aria & Cuccurullo, 2017) y Gephi (Bastian et al., 2009). Nota: **mismo valor.**

Una exploración mediante un test Wilcoxon de suma de rangos mostró que no hay diferencias significativas en la mediana del grado de las instituciones de ALC o China (i.e., mediana de grado por región: 1997-2001 – ALC: 99.5, China: 96.5; 2002-2007 – ALC: 22, China: 5; 2008-2013 – ALC: 7, China: 7; 2014-2019 – ALC: 105, China: 91). Luego, ni LAC o China están tomando la delantera o quedándose atrás comparando la mediana del número total de coautorías interinstitucionales (Figura 2).

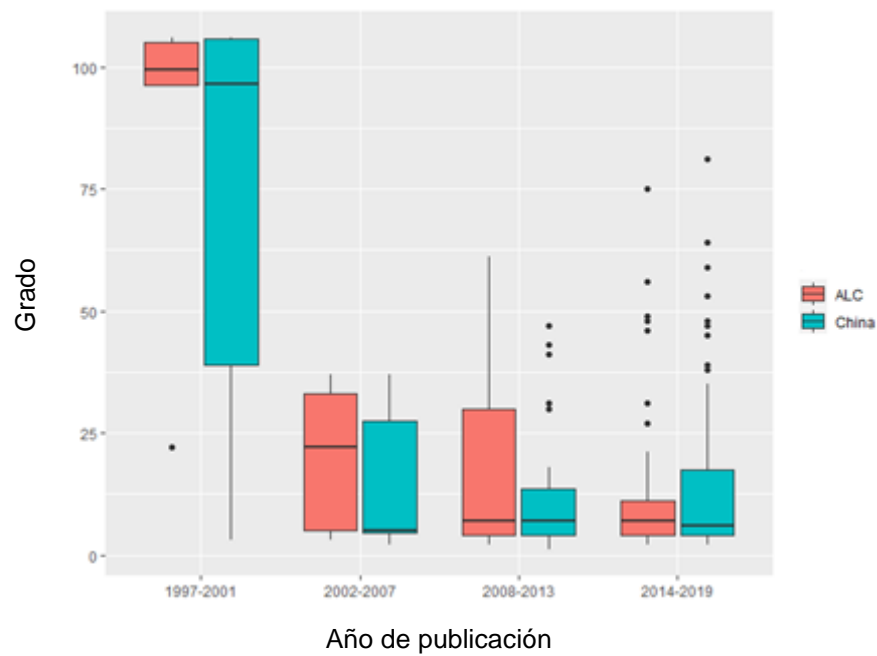

**Figura 2 Diagrama de cajas - grados de cada institución por periodo y región. Fuente: el autor con base en Scopus (2019) y procesado con bibliometrix (Aria & Cuccurullo, 2017).**

Los puentes interinstitucionales (i.e., los de mayor intermediación) se encuentran localizados en latitudes distintas a LAC o China, como Europa y, principalmente, Australia. Ahora bien, entre ALC y China, este último sobresale. En el caso latinoamericano, U. Brasilia, Brasil en 1997-2001 (Figura 3a – nodo 91) y U. Andes, Colombia en 2014-2019 (Figura 3a – nodo 31) fueron las únicas instituciones que figuraron en alguno de los top-5 de mayor intermediación. En el caso chino, figuran Beijing U. en 1997-2001 (Figura 3a – nodo 12) y Lignan U. (Figura 3a – nodo 5) y Capital Normal U. (Figura 3a – nodo 3) en 2002-2007.

Los top-5 restantes fueron complementados por: Babeş-Bolyai U., Rumania (Figura 3a-b – nodo 2), Sultan Qaboos U., Omán (Figura 3a-b – nodo 6), U. Ljubljana, Eslovenia (Figura 2a-b – nodos 12 y 7), U. Philippines, Filipinas (Figura 3c – nodo 7), Victoria U. Wellington, Australia (Figura 3c – nodo 8), U. California, EEUU (Figura 3c – nodo 5), U. Texas, EEUU (Figura 3c – nodo 36), Monash U., Australia (Figura 3c – nodo 20), Aarhus U., Dinamarca (Figura 3d – nodo 23), U. New South Wales, Australia (Figura 3d – nodo 102), U. Gothenburg, Suecia (Figura 3d – nodo 20), y U. Queensland, Australia (Figura 3d – nodo 39).

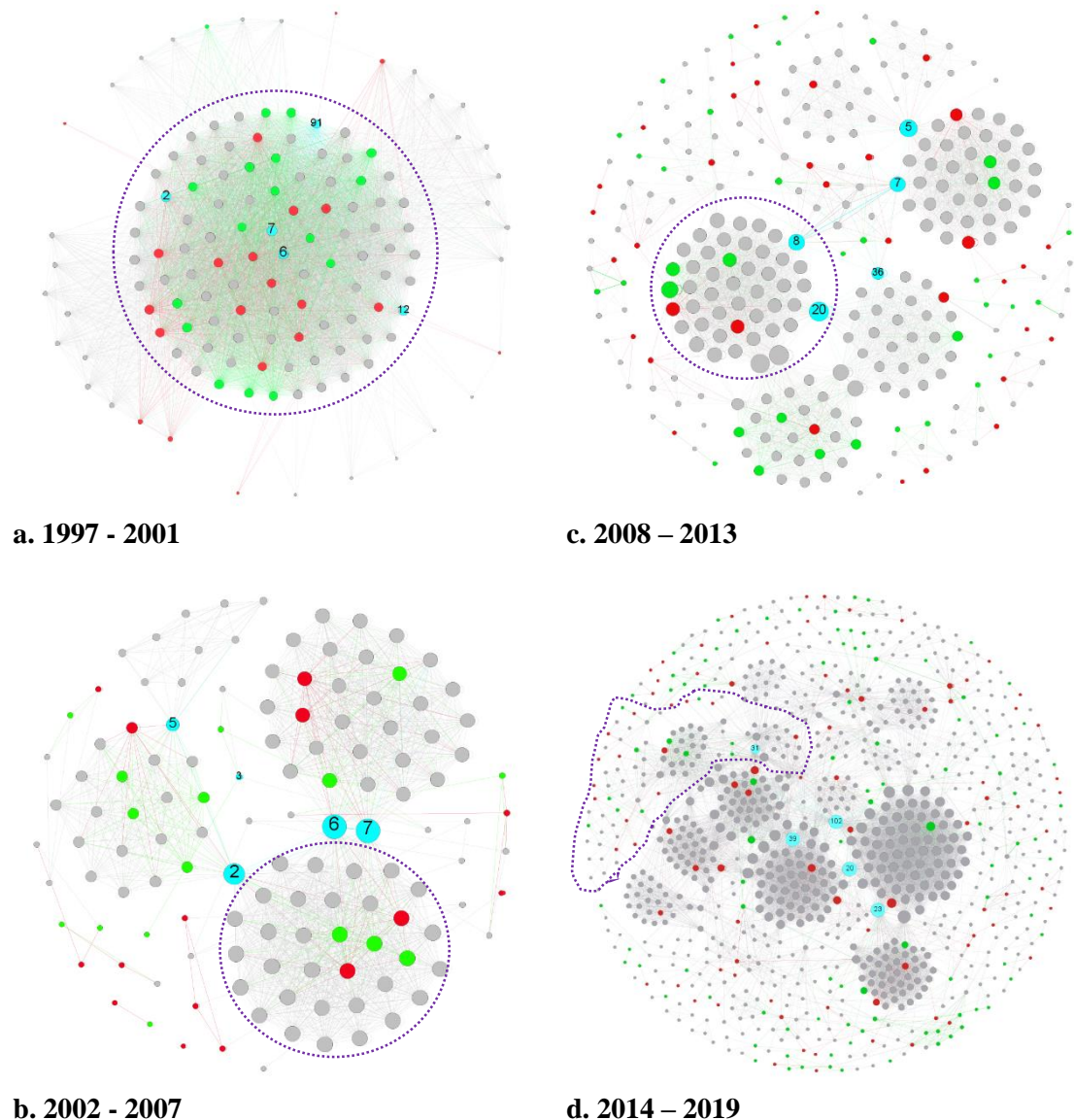

**a. 1997 - 2001**    **c. 2008 – 2013**

**b. 2002 - 2007**    **d. 2014 – 2019**

**Figura 3 Redes de coautoría LAC-China 1997-2019. Fuente: el autor con base en Scopus (2019) y procesado con bibliometrix (Aria & Cuccurullo, 2017) y Gephi (Bastian et al., 2009). Nota: el tamaño de los nodos es proporcional al grado. Los nodos de color verde representan las instituciones de ALC, los de color rojo las de China, y los de color gris el resto del mundo. Los nodos de color azul aguamarina representan los nodos del top-5 con mayor intermediación en cada periodo. La figura punteada de color púrpura encierra el componente principal.**

El incremento de la productividad de ambas regiones, y del mundo en general (Bornmann & Mutz, 2015), especialmente en el último sexenio, es un reflejo del avance de China como líder en la inversión y productividad en actividades de I+D (Tollefson, 2018). Por ejemplo, Quan et al. (2017) sostienen que universidades en China ofrecen incentivos que van desde USD$30 a USD$165,000 por artículo publicado en WoS. Con todo, aún se debate con intensidad la relación directa y positiva entre las políticas de incentivos por publicaciones, su calidad, futuro impacto en la investigación y los efectos adversos sobre la salud mental de los futuros investigadores debido las exigencias institucionales en materia de productividad y eficiencias (Cortés-Sánchez, 2019, 2020a; Cyranoski, 2018; Dell'Anno et al., 2020; Levecque et al., 2017; Lindner et al., 2018; Macháček & Srholec, 2021; Smaldino & McElreath, 2016).

Llama la atención las métricas macro y micro del período 1997-2001 durante el cual se publicaron solamente cuatro artículos. En efecto, en 1999 se publicó un artículo excepcional en el cual participan más de 50 autores de todo el mundo como parte del programa de investigación GLOBE (*Global Leadership and Organizational Behavior Effectiveness*) el cual tuvo ejecución en más de 60 *culturas* (Den Hartog et al., 1999). La participación de investigadores de ambas regiones en el mismo estudio explica la similitud en su punto de partida (i.e., ubicación central en la red, grado e intermediación). La comunidad compacta de instituciones de ALC y China en dicho periodo, se fue atomizando: las instituciones latinoamericanas fueron ocupando la periferia, mientras las chinas el centro.

La densidad de redes de coautoría en negocios-administración reportada en otras investigaciones, es significativamente menor (i.e., 0.0002) (Acedo et al., 2006) aunque similar en economía (i.e., 0.013) (de Stefano et al., 2011). Así, las redes de ALC y China aquí modeladas son más densas. En todo caso, se debe considerar que las redes de coautoría institucional tienen menos nodos que las de investigadores individuales empleadas para esta comparación.

El incremento de la modularidad, revela que, como se dice en Colombia, se han formado *roscas*: comunidades macizas lánguidamente enlazadas con comunidades externas. Esta dinámica puede ser entendida a través de dos postulados: que las redes se expanden continuamente agregando nuevos nodos y que los nuevos nodos no se conectan aleatoriamente con cualquier nodo ya que buscan enlazarse con nodos elevadamente conectados (Barabási & Albert, 1999).

Aunque no se identificaron diferencias significativas en la mediana del grado de los nodos en ambas regiones, las instituciones de China han cobrado un papel como puentes sociales hasta cierto periodo (2002-2007), lo que ha sido señalado anteriormente (Finardi & Buratti, 2016). Se debe resaltar que, durante el último sexenio, U. Andes, Colombia, figuró en el top-5 de las universidades con mayor intermediación entre ambas regiones, especialmente al interior de ALC (Cortés-Sánchez, 2020b). Sin previsión alguna, se debe considerar el papel de las universidades australianas como geografía que concentra la mayor cantidad de instituciones con mayor intermediación. Un categórico resultado de la reforma económica de China (1978) y otros factores como la migración o acuerdos interinstitucionales entre China y Australia (Hugo, 2010; Niu, 2014).

## 4   Conclusión

Los resultados sobre la colaboración científica entre ALC y China durante 1997-2019 en negocios y economía, mostraron que la coautoría entre ambas regiones fue fertilizada mediante su participación en proyectos globales relacionados con liderazgo, para, después divergir: ALC hacia la periferia, China hacia el centro. Así, se comienzan a consolidar los resultados de la política de inversión e incentivos nacional/institucional de China. El impacto de su calidad en el largo plazo sigue siendo un interrogante para futuras investigaciones. La incursión de nuevas instituciones en la red, va direccionada a instituciones o comunidades altamente conectadas, confirmando cúmulos compactos con enlaces reducidos hacia otras comunidades. Frente a los puentes sociales, China exhibió su protagonismo, que, a la postre, se fue desvaneciendo para dar lugar a una institución colombiana. Surge la intermediación de múltiples instituciones australianas, lo que sugiere que en esa geografía se concentran los puentes sociales entre ALC y China y el Norte Global. Futuras indagaciones que respondan a las limitaciones de este estudio, podrían ahondar en una evaluación del impacto del tsunami de publicaciones de China, emplear bases de datos bibliográficas diferentes a Scopus (e.g., WoS, Google Scholar, Dimensions), implementar técnicas de cartografía científica no redundantes (e.g., acoplamiento bibliográfico, co-ocurrencia de palabras, entre otros), y refinar el análisis a coautores en lugar de instituciones.

## Referencias

Acedo, F. J., Barroso, C., Casanueva, C., & Galán, J. L. (2006). Co-authorship in management and organizational studies: An empirical and network analysis. *Journal of Management*

*Studies*, *43*(5), 957–983. https://doi.org/10.1111/j.1467-6486.2006.00625.x

Adams, J. (2012). The rise of research networks. *Nature*, *490*(7420), 335–336. https://doi.org/10.1038/490335a

Adams, J., Pendlebury, D., Porter, R., & Szomszor, M. (2019). *Global Research Report - Multi-authorship and research analytics*.

Aria, M., & Cuccurullo, C. (2017). bibliometrix: An R-tool for comprehensive science mapping analysis. *Journal of Informetrics*, *11*(4), 959–975. https://doi.org/10.1016/j.joi.2017.08.007

Baas, J., Schotten, M., Plume, A., Côté, G., & Karimi, R. (2020). Scopus as a curated, high-quality bibliometric data source for academic research in quantitative science studies. *Quantitative Science Studies*, *1*(1), 377–386. https://doi.org/10.1162/qss_a_00019

Barabási, A. L., & Albert, R. (1999). Emergence of scaling in random networks. *Science*, *286*(5439), 509–512. https://doi.org/10.1126/science.286.5439.509

Bastian, M., Heymann, S., & Jacomy, M. (2009). Gephi: an open source software for exploring and manipulating networks. *International AAAI Conference on Weblogs and Social Media*. https://gephi.org/users/publications/

Blondel, V. D., Guillaume, J.-L., Lambiotte, R., & Lefebvre, E. (2008). Fast unfolding of communities in large networks. *Journal of Statistical Mechanics: Theory and Experiment*, *2008*(10), P10008.

Bornmann, L., & Mutz, R. (2015). Growth rates of modern science: A bibliometric analysis based on the number of publications and cited references. *Journal of the Association for Information Science and Technology*, *66*(11), 2215–2222. https://doi.org/10.1002/asi.23329

Brandes, U. (2001). A faster algorithm for betweenness centrality. *Journal of Mathematical Sociology*, *25*(2), 163–177. https://doi.org/10.1080/0022250X.2001.9990249

CEPAL. (2012). *China y América Latina y el Caribe: hacia una relación económica y comercial estratégica*. https://www.cepal.org/es/publicaciones/2598-china-america-latina-caribe-relacion-economica-comercial-estrategica

CEPAL. (2019). *Panorama Social de América Latina 2019*. https://repositorio.cepal.org/handle/11362/44969

Connell, M., & Cornejo Bustamante, R. (1992). *China - América Latina: Génesis y desarrollo de sus relaciones* (1ra ed.). El Colegio de México.

Cortés-Sánchez, J. D. (2019). Innovation in Latin America through the lens of bibliometrics: crammed and fading away. *Scientometrics*, *121*(2), 869–895. https://doi.org/10.1007/s11192-019-03201-0

Cortés-Sánchez, J. D. (2020a). A bibliometric outlook of the most cited documents in business, management and accounting in Ibero-America. *European Research on Management and Business Economics*, *26*(1), 1–8. https://doi.org/10.1016/j.iedeen.2019.12.003

Cortés-Sánchez, J. D. (2020b). *Atlas de la investigación en administración en América Latina Vol. 4* (Núm. 158). https://repository.urosario.edu.co/handle/10336/30108

Cortés-Sánchez, J. D. (2022). Research on Innovation in China and Latin America: Bibliometric Insights in the Field of Business, Management and Decision Sciences. *Latin American Business Review*, *23*(2).

Cyranoski, D. (2018). China awaits controversial blacklist of "poor quality" journals. *Nature*, *562*(7728), 471–472. https://doi.org/10.1038/d41586-018-07025-5

de Stefano, D., Giordano, G., & Vitale, M. P. (2011). Issues in the analysis of co-authorship networks. *Quality and Quantity*, *45*(5), 1091–1107. https://doi.org/10.1007/s11135-011-9493-2

Dell'Anno, R., Caferra, R., & Morone, A. (2020). A "Trojan Horse" in the peer-review process of fee-charging economic journals. *Journal of Informetrics*, *14*(3), 101052. https://doi.org/10.1016/j.joi.2020.101052

Den Hartog, D. N., House, R. J., Hanges, P. J., Antonio Ruiz-Quintanilla, S., Dorfman, P. W., Abdalla, I. A., Adetoun, B. S., Aditya, R. N., Agourram, H., Akande, A., Akande, B. E., Akerblom, S., Altschul, C., Alvarez-Backus, E., Andrews, J., Arias, M. E., Arif, M. S., Ashkanasy, N. M., Asllani, A., … Zhou, J. (1999). Culture specific and crossculturally generalizable implicit leadership theories: Are attributes of charismatic/transformational leadership universally endorsed? *Leadership Quarterly*, *10*(2), 219–256.

https://doi.org/10.1016/S1048-9843(99)00018-1

Desrochers, N., Paul-Hus, A., & Larivière, V. (2016). The Angle Sum Theory: Exploring the literature on Acknowledgments in Scholarly Communication. En C. R. Sugimoto (Ed.), *Theories of Informetrics and Scholarly Communication* (pp. 225–246). De Gruyter Saur.

Finardi, U., & Buratti, A. (2016). Scientific collaboration framework of BRICS countries: an analysis of international coauthorship. *Scientometrics*, *109*(1), 433–446. https://doi.org/10.1007/s11192-016-1927-0

FLAUC. (s/f). *Fudan Latin America University Consortium - FLAUC*. Recuperado el 1 de marzo de 2021, de https://flauc.fudan.edu.cn/

Gingras, Y. (2014). *Bibliometrics and research evaluation - Uses and abuses*. The MIT Press.

Gonzalez-Vicente, R. (2012). Mapping Chinese Mining Investment in Latin America: Politics or Market? *The China Quarterly*, *209*, 35–58. http://www.jstor.org/stable/41447821

Hugo, G. (2010). The Indian and Chinese Academic Diaspora in Australia: A Comparison. *Asian and Pacific Migration Journal*, *19*(1), 87–116. https://doi.org/10.1177/011719681001900105

Iacobucci, D., McBride, R., Popovich, D. L., & Rouziou, M. (2018). In Social Network Analysis, Which Centrality Index Should I Use?: Theoretical Differences and Empirical Similarities among Top Centralities. *Journal of Methods and Measurement in the Social Sciences*, *8*(2), 72–99. https://doi.org/10.2458/v8i2.22991

Ioannidis, J. P. A., Klavans, R., & Boyack, K. W. (2016). Multiple Citation Indicators and Their Composite across Scientific Disciplines. *PLOS Biology*, *14*(7), 1–17. https://doi.org/10.1371/journal.pbio.1002501

Lemarchand, G. A. (2012). The long-term dynamics of co-authorship scientific networks: Iberoamerican countries (1973-2010). *Research Policy*, *41*(2), 291–305. https://doi.org/10.1016/j.respol.2011.10.009

Levecque, K., Anseel, F., De Beuckelaer, A., Van der Heyden, J., & Gisle, L. (2017). Work organization and mental health problems in PhD students. *Research Policy*, *46*(4), 868–879. https://doi.org/10.1016/j.respol.2017.02.008

Li, E. Y., Liao, C. H., & Yen, H. R. (2013). Co-authorship networks and research impact: A social capital perspective. *Research Policy*, *42*(9), 1515–1530. https://doi.org/10.1016/j.respol.2013.06.012

Lin, J. Y., & Treichel, V. (2012). *Learning from China's Rise to Escape the Middle-Income Trap: A New Structural Economics Approach to Latin America*. The World Bank. https://doi.org/10.1596/1813-9450-6165

Lindner, M. D., Torralba, K. D., & Khan, N. A. (2018). Scientific productivity: An exploratory study of metrics and incentives. *PLOS ONE*, *13*(4), e0195321. https://doi.org/10.1371/journal.pone.0195321

Macháček, V., & Srholec, M. (2021). Predatory publishing in Scopus: evidence on cross-country differences. *Scientometrics*, 1–25. https://doi.org/10.1007/s11192-020-03852-4

Mesquita Moreira, M. (2007). Fear of China: Is There a Future for Manufacturing in Latin America? *World Development*, *35*(3), 355–376. https://doi.org/10.1016/j.worlddev.2006.11.001

Mongeon, P., & Paul-Hus, A. (2016). The journal coverage of Web of Science and Scopus: a comparative analysis. *Scientometrics*, *106*(1), 213–228. https://doi.org/10.1007/s11192-015-1765-5

Myers, M., & Wise, C. (Eds.). (2016). *The Political Economy of China-Latin America Relations in the New Millennium: Brave New World*. Taylor & Francis.

Niu, X. S. (2014). International scientific collaboration between Australia and China: A mixed-methodology for investigating the social processes and its implications for national innovation systems. *Technological Forecasting and Social Change*, *85*, 58–68. https://doi.org/10.1016/j.techfore.2013.10.014

Opsahl, T., Agneessens, F., & Skvoretz, J. (2010). Node centrality in weighted networks: Generalizing degree and shortest paths. *Social Networks*, *32*(3), 245–251. https://doi.org/10.1016/j.socnet.2010.03.006

Quan, W., Chen, B., & Shu, F. (2017). Publish or impoverish: An investigation of the monetary

reward system of science in China (1999-2016). *Aslib Journal of Information Management*, *69*(5), 486–502. https://doi.org/10.1108/AJIM-01-2017-0014

R Core Team. (2014). R: A language and environment for statistical computing. En *R Foundation for Statistical Computing, Vienna, Austria.* (Vol. 0, pp. 1–2667). R Foundation for Statistical Computing. online: http://www. R-project. org

Rohde, R. A., & Muller, R. A. (2015). Air Pollution in China: Mapping of Concentrations and Sources. *PLOS ONE*, *10*(8), e0135749. https://doi.org/10.1371/journal.pone.0135749

Scopus. (2019). Scopus. En *How Scopus works*. https://www.elsevier.com/solutions/scopus/how-scopus-works

Scopus. (2020). *Scopus - Document search*. https://www.scopus.com/search/form.uri?display=basic

Scott, J. (1988). Social Network Analysis. *Sociology*, *22*(1), 109–127. https://doi.org/10.1177/0038038588022001007

Smaldino, P. E., & McElreath, R. (2016). The natural selection of bad science. *Royal Society Open Science*, *3*(9). https://doi.org/10.1098/rsos.160384

Tollefson, J. (2018). China declared world's largest producer of scientific articles. *Nature*, *553*(7689), 390. https://doi.org/10.1038/d41586-018-00927-4

Traag, V., Waltman, L., & Eck, N. J. (2019). From Louvain to Leiden: guaranteeing well-connected communities. *Scientific Reports*, *9*.

Wagner, C. S., Park, H. W., & Leydesdorff, L. (2015). The Continuing Growth of Global Cooperation Networks in Research: A Conundrum for National Governments. *PLOS ONE*, *10*(7), e0131816. https://doi.org/10.1371/journal.pone.0131816

Wuchty, S., Jones, B. F., & Uzzi, B. (2007). The Increasing Dominance of Teams in Production of Knowledge. *Science*, *316*(5827), 1036–1039. https://doi.org/10.1126/science.1136099